\documentclass{elsart4}

\usepackage{graphics}
\usepackage{graphicx}
\usepackage{epsfig}
\usepackage{amssymb}
\journal{Physica B}

\def\bscco{Bi$_2$Sr$_2$CaCu$_2$O$_{8+\delta}$}

\def\lco{La$_2$CuO$_4$}

\def\lsno{La$_{2-x}$Sr$_x$NiO$_4$}
\def\lsco{La$_{2-x}$Sr$_x$CuO$_4$}
\def\lbco{La$_{2-x}$Ba$_x$CuO$_4$}
\def\lbcoate{La$_{1.875}$Ba$_{0.125}$CuO$_4$}
\def\lnsco{La$_{1.6-x}$Nd$_{0.4}$Sr$_x$CuO$_4$}
\def\lesco{La$_{1.8-x}$Eu$_{0.2}$Sr$_x$CuO$_4$}
\def\ybco{YBa$_2$Cu$_3$O$_{6+x}$}

\def\bscco{Bi$_2$Sr$_2$CaCu$_2$O$_{8+\delta}$}

\def\tco{$T_{\rm co}$}

\begin{document}

\begin{frontmatter}


\title{Stripes and Superconductivity in Cuprates}
\author{John M. Tranquada\corauthref{cor1}}
\address{Condensed Matter Physics \&\ Materials Science Dept., Brookhaven National Laboratory, Upton, NY 11973-5000, USA}
 \corauth[cor1]{e-mail: jtran@bnl.gov}





\begin{abstract}
Holes doped into the CuO$_2$ planes of cuprate parent compounds frustrate the antiferromagnetic order.  The development of spin and charge stripes provides a compromise between the competing magnetic and kinetic energies.  Static stripe order has been observed only in certain particular compounds, but there are signatures which suggest that dynamic stripe correlations are common in the cuprates.  Though stripe order is bad for superconducting phase coherence, stripes are compatible with strong pairing.  Ironically, magnetic-field-induced stripe order appears to enhance the stability of superconducting order within the planes.
\end{abstract}

\end{frontmatter}

\section{Introduction}
\label{}

A quarter century after the discovery of high-temperature superconductivity \cite{bedn86}, the nature of hole-doped copper-oxide compounds remains controversial.  The theoretical machinery that has been developed to describe conventional superconductors is built on top of Fermi liquid theory, but there is considerable experimental evidence suggesting that the normal state properties of cuprates are inconsistent with Fermi liquid predictions over much of the interesting part of the phase diagram.  The usual starting point of band theory, in which minimizing the kinetic energy of the conduction electrons plays a dominant role, is inadequate.  Electron-electron interactions play a crucial role; however, they are not so strong that one can apply a perturbation theory about the strong-coupling limit.  Interactions and kinetic energy are roughly comparable at the doping levels where superconductivity occurs, and this intermediate-coupling regime poses a particular challenge to theory.

Experiments indicate that the doped cuprates have a mixture of characters.  To see this, we start with an undoped parent compound such as \lco.  This material is an insulator with a charge-transfer gap of $\sim2$~eV.  Antiferromagnetism develops within the CuO$_2$ planes as a consequence of strong onsite Coulomb repulsion between electrons in the same Cu $3d_{x^2-y^2}$ orbital \cite{kast98}.  The effective magnetic interaction is well characterized by the superexchange mechanism \cite{ande87}, and the magnetic excitation spectrum is described quite well by spin-wave theory with nearest-neighbor superexchange energy $J\sim 140$~meV \cite{head10}.

Things start to change as soon one begins to dope holes into the planes.  In \lsco, the long-range antiferromagnetic (AF) order is destroyed by $x=0.02$, to be replaced by a spin-glass phase with incommensurate magnetic order that develops below 10~K; furthermore, there is evidence for phase separation between the AF and incommensurate phases for $x<0.02$ \cite{mats02}.  The rapid destruction of AF order as mobile charge carriers are introduced indicates a competition between the tendency of the holes to delocalize, in order to reduce their kinetic energy, and the onsite Coulomb interactions that drive the AF correlations.  Part of this effect can be understood if we think about a single-band model (Cu-sites only) and consider an individual hole moving in an AF background \cite{trug88,lau11}.  An electron near the Fermi energy moving in the so-called ``nodal'' direction can hop on the same AF sublattice; no spin flips are involved, so there is no conflict with the AF correlations.  In contrast, an electron hopping in the antinodal direction, along Cu-O bonds, can only do so by flipping spins and disrupting the AF correlations.   The impact of the doped holes on the AF background becomes even more obvious when one takes account of the fact that doped holes have strong O $2p$ character, as a hole localized between on an O between a pair of Cu ions will induce a ferromagnetic alignment of those Cu spins \cite{emer88,ahar88}.  In a different direction, there have been proposals that doping should simply cause the spin correlations to change to a spiral form \cite{lusc07}; however, in that case, one would not expect to see such a dramatic loss of order at $x=0.02$.   

Neutron scattering experiments show that dynamical AF correlations survive in the doped cuprates throughout most of the superconducting range \cite{fuji11},  but how can a symptom of the correlated insulator state coexist with itinerant charge carriers?

\section{Stripe order}

One way for locally-AF spin correlations to coexist with mobile holes is through the formation of charge and spin stripes.  These stripe patterns are easiest to analyze when they are statically ordered.  Theoretical motivations for stripes have been reviewed in \cite{kive03,vojt09,zaan01}.  Experimentally, charge and spin stripe order is actually quite common in layered, transition-metal-oxide compounds such as \lsno\ \cite{yosh00} and La$_{2-x}$Sr$_x$CoO$_4$ \cite{saki08,cwik09}.  Of course, these latter systems tend to be insulating when stripe ordered.  A key difference in the cuprates is that magnetic Cu ions have a single unpaired $3d$ electron, with spin $S=1/2$, which is essential to the mobility of the doped holes \cite{tran98c}.

The first indication of an anomaly possibly associated with stripes was the discovery of a sharp depression in the superconducting transition temperature, $T_c$, centered on the doping level $x=1/8$ in \lbco\ (LBCO) \cite{mood88}, an effect not observed (or, at least, not as strongly) in \lsco\ (LSCO).  An x-ray diffraction study by Axe {\it et al.} \cite{axe89} demonstrated that \lbco\ exhibits a subtle structural phase transition at low temperatures that causes orthogonal Cu-O bonds within each plane to become inequivalent.  (The orientation of the anisotropy rotates $90^\circ$ from layer to layer.)  This structural anisotropy appears to be important to the development of static stripe order, which was first detected by neutron diffraction in the isostructural compound La$_{1.48}$Nd$_{0.4}$Sr$_{0.12}$CuO$_4$ \cite{tran95a}.   It proved challenging to grow \lbco\ crystals at the same hole concentration, but eventually Fujita and collaborators \cite{fuji04} were successful, allowing confirmation of stripe order in \lbcoate.  

The phase diagram for charge and spin stripe order in \lbco\ has now been established as a function of doping through a combination of neutron and x-ray diffraction, as well as magnetic susceptibility measurements \cite{huck11}.  While the amplitude of the stripe order is greatest at $x=1/8$, weak charge stripe order is still detectable at $x=0.095$ and 0.155.  Measurements by other groups \cite{duns08,adac01} are consistent with the phase diagram after one calibrates the relative compositions through the doping dependence of structural phase transition temperatures \cite{huck11}.  The maximum amplitude of the stripe order is correlated with a minimum of $T_c$ for bulk superconductivity; however, there is evidence of superconducting correlations at much higher temperatures \cite{li07,tran08}, as will be discussed.

An important question concerns the strength of the stripe order.  Does it involve substantial local magnetic moments or a weak spin-density modulation?  How big is the charge modulation?  One measure of the moment size is given by muon spin rotation measurements, which probe the local hyperfine field at the muon site.  Analysis of such measurements indicates a maximum ordered moment at low temperature of $\sim0.3$~$\mu_{\rm B}$, about 60\%\ of the value in AF La$_2$CuO$_4$ \cite{nach98}; this is a substantial value, considering the importance of quantum fluctuations.  There have also been measurements of the anisotropic bulk susceptibility on a single crystal \cite{huck08}.  Below the spin-stripe-ordering transition, a large temperature-dependent anisotropy of the susceptibility is found, consistent with what one would expect to see in an AF insulator.  With the field aligned along a Cu-O bond direction, a spin-flop transition is observed at $\sim6$~T \cite{huck08}, again consistent with behavior typically found in systems where a local-moment description is appropriate.

Regarding the charge, neutron and hard-x-ray diffraction are sensitive just to atomic displacements, which provide only an indirect measure of charge modulation.  Abbamonte {\it et al.}\cite{abba05} used soft x-ray scattering to demonstrate that the charge-order diffraction peak is resonant at the energy of the O $2p$ pre-edge peak in the O $K$-edge absorption spectrum.  This directly demonstrates that the occupancy of the O $2p$ states is spatially modulated with the period of the charge stripes.  Quantitative analysis indicated a substantial modulation amplitude \cite{abba05}, consistent with a mean-field calculation by Lorenzana and Seibold \cite{lore02}.  Such resonant scattering measurements have since been used to determine the charge-ordering phase diagram for \lesco \cite{fink11}, and to confirm the charge-stripe order in La$_{1.48}$Nd$_{0.4}$Sr$_{0.12}$CuO$_4$ \cite{wilk11}.

Weak incommensurate spin order has also been observed in rather underdoped \ybco\ \cite{haug09}, and this order is enhanced by an applied magnetic-field.  At slightly larger hole doping, similar order can be induced by substituting Zn for 2\%\ of the Cu atoms \cite{such10}.  While the
observed incommensurability at a given hole concentration is systematically smaller than that observed in \lbco\ \cite{haug10}, the trend with doping and the orientation of the modulation wave vector is quite similar.  Very recently, Wu {\it et al.} \cite{wu11} have reported evidence from nuclear magnetic resonance measurements for charge stripe order at a hole concentration near 1/8 induced by magnetic fields greater than 25~T applied perpendicular to the planes.

\section{Stripe phase is 2D}

While stripe order involves a unidirectional modulation, the electronic character of the stripe-ordered layer is two-dimensional (2D).  Model calculations \cite{gran08} indicate that the Fermi surface corresponding to dispersion along the charge stripes consists of flat sections in the antinodal $(\pi,0)$ and $(0,\pi)$ regions of reciprocal space, while the states along the Fermi arc, extending about $(\pi/2,\pi/2)$, are much more homogeneous in terms of distribution in real space.   It has been pointed out that the Fermi arc states have dominant oxygen character, while the antinodal states have more copper character \cite{niks11}.

Experimental angle-resolved photoemission spectroscopic (ARPES) studies on stripe-ordered\linebreak 
\lbcoate\ display behavior very similar to that seen in other cuprate superconductors \cite{vall06,he09}.  The spectral function measured vs.\ energy shows sharp peaks along the Fermi arc, but much broader features in the antinodal region.  A $d$-wave like gap is found along the Fermi arc for $T\lesssim 40$~K, while a somewhat larger gap is present in the antinodal states.  The Fermi arc and antinodal gap have also been observed in stripe-ordered La$_{1.48}$Nd$_{0.4}$Sr$_{0.12}$CuO$_4$ \cite{chan08b}.  Despite the coexisting magnetic order, there are no obvious features in the ARPES spectra of either of these systems that would indicate the presence of stripe order.  

Transport properties are also consistent with the stripe-ordered state remaining effectively metallic.  The in-plane resistivity is dominated by states near the Fermi arc, and in \lbcoate\ it retains a metallic temperature derivative below the charge-ordering temperature \tco\ \cite{li07}.  It should be noted that Adachi {\it et al.} \cite{adac11} have seen an upturn in the in-plane resistivity on cooling below \tco; however, the measurements are challenging due to the extreme anisotropy of the electronic properties \cite{wen11}.  Optical conductivity measurements with in-plane polarization show a narrowing of the Drude peak just below \tco, consistent with metallic behavior \cite{home06,home11}.   Low-temperature metallic behavior is not universal for all stripe-ordered cuprates.  For example, significant upturns in the in-plane resistivity have been observed in \lnsco\ \cite{ichi00,daou09b} and there is an absorption peak in the low-frequency optical conductivity for La$_{1.275}$Nd$_{0.6}$Sr$_{0.125}$CuO$_4$ \cite{dumm02}.

\section{Diffraction vs.\ scanning tunneling spectroscopy}

Real-space modulations of tunneling conductance have been imaged on cleaved samples of \bscco\ (Bi2212) and
\linebreak
Bi$_{2-y}$Pb$_y$Sr$_{2-z}$La$_z$CuO$_{6+x}$ (Bi2201) \cite{howa03b,kohs07,lawl10,park10,wise08}.  The modulations have a period of $\sim4$ lattice spacings, which is similar to the charge stripe period in the $n_h=1/8$ phase; however, the STS modulation wave vector decreases with doping \cite{wise08}, whereas the stripe wave vector increases with doping, before saturating above $n_h=1/8$ \cite{birg06}.  Now, these results have been measured in different systems, so they are not in direct conflict; however, unpublished data on Bi2201 from Fujita, Enoki, and Yamada indicate that the doping dependence of the spin-stripe wave vector is actually quite similar to that in LSCO.  This leads to a challenge in how to reconcile the results of these distinct techniques.

It has been proposed, based on ARPES work, that the modulation seen by STS corresponds to $2k_{\rm F}$ scattering determined by the parallel Fermi surface sheets in the antinodal regions \cite{shen05}.  This matches fairly well in magnitude and doping dependence.  On the other hand, in a recent ARPES study \cite{gweo11}, it has been argued that there is a small but systematic discrepancy between these quantities.  It should be noted that the ``stripe''-like modulation signal in STS is strongest at a substantial bias voltage, on the order of the pseudogap energy, and the relevant wave vector at that energy might be shifted from that at the Fermi energy (zero bias voltage).  Furthermore, the pseudogap energy is measured from these antinodal states, so it seems likely that they are connected.

If STS is detecting a $2k_{\rm F}$-like modulation, it is not incompatible with stripes.  As discussed above, charge stripes should have Fermi surface segments in the antinodal regime \cite{gran08}; however, the associated modulation would be orthogonal to the charge-stripe modulation.  Parker {\it et al.} \cite{park10} have shown in Bi2212 that the STS modulation strength and temperature onset maxima occur close to $n_h=1/8$, supporting a connection to charge stripes.

\section{Stripe dynamics}

Spin fluctuations disperse from the incommensurate magnetic superlattice peaks in the stripe ordered phase \cite{fuji04,duns08,tran99a}.  In \lbcoate, the spectrum has been measured up to $\sim200$~meV \cite{tran04}, and it is found to share the same ``hour-glass" dispersion as other cuprate superconductors \cite{fuji11}.  The energy $E_{\rm cross}$ of the crossing point of the spectrum varies linearly with doping on the underdoped side, decreasing towards zero as the doping is reduced.  Thus, the upwardly dispersing excitations evolve into the spin waves of the AF phase.  The spectrum below $E_{\rm cross}$ is strongly modified from the AF behavior because of the presence of the doped holes.  Many researchers have attempted to explain the downwardly dispersing spectrum in terms of particle-hole excitations of a homogeneous system \cite{esch06}; however, such a mechanism is challenged when it comes to explaining a number of the features observed in LBCO.  

In \lbcoate, the low-energy incommensurate spin fluctuations survive above \tco, indicating the presence of fluctuating stripes \cite{xu07}.  These low-energy excitations are very similar to those in LSCO throughout the underdoped regime, suggesting that fluctuating stripes are a common feature.

At modest energies, the magnetic spectral weight in the doped cuprates is comparable to that of AF spin waves, but Stock {\it et al.} \cite{stoc10} have pointed out that the spectral weight is suppressed above the pseudogap energy.  This observation suggests that particle-hole excitations compete with spin fluctuations, rather than reinforcing them.  Stripe-like segregation of spins and holes provides a mechanism to maintain AF correlations with local-moment character.  There must be a balance between the AF energy and the kinetic energy of the holes, so it is reasonable to expect that the stripe-like correlations will eventually disappear at sufficient hole density.

\section{Stripes and superconductivity}

While stripe order competes with bulk superconducting order, evidence has been found for strong two-dimensional superconducting (SC) correlations that begin at a temperature comparable to the typical bulk $T_c$ \cite{li07,tran08}.  It is quite unusual to observe 2D SC in a bulk crystal, as interlayer Josephson coupling inevitably leads to 3D SC order.  To explain the decoupling, a pair density wave (PDW) SC state has been proposed \cite{hime02,berg07}.  In the PDW state, the pair wave function is locally $d$-wave-like, but it is modulated by an envelope function that oscillates sinusoidally, with extrema aligned with the charge stripes, and nodes centered in the spin stripes.

Berg {\it et al.} \cite{berg09a} considered mechanisms that would induce the antiphase coupling of neighboring SC stripes.  A couple of recent calculations based on the 2D $t$-$J$ model have identified conditions where the PDW phase appears to be the ground state.  Loder {\it et al.} \cite{lode11} have analyzed mean field models valid for either small or large $J/t$, while Corboz {\it et al.} \cite{corb11} made use of a new type of variational Monte Carlo scheme for an intermediate value of $J/t$.  In the latter calculation, there is no significant difference in energy between in-phase and antiphase coupling between superconducting stripes.  

Although stripe order can be bad for SC phase coherence, it appears to be compatible with strong pairing.  Kivelson and Fradkin \cite{kive07} have argued that stripe-like inhomogeneity can enhance pairing and superconductivity.  Empirically, a close relationship between stripes and superconductivity is suggested by the fact that stripe ordering temperatures are always close to $T_c$ values.  Dynamic fluctuations of the inhomogeneity may be essential for optimizing the phase order at high temperature.

\section{Magnetic-field-induced stripe order}

It has been known for some time that spin stripe order can be induced by a $c$-axis magnetic field \cite{lake02}.  Recent measurements of LBCO with $x=0.095$, which has only very weak stripe order in zero field \cite{huck11}, have shown for the first time that charge stripe order can also be enhanced by a field \cite{wen11}.  The presence of the stripe order appears to weaken the interplanar Josephson coupling, but does not reduce it to zero.  Transport measurements indicate that the $c$-axis field can can cause the interlayer resistivity to become finite while the in-plane resistivity remains zero \cite{wen11}.  (Such behavior violates conventional theoretical expectations, as one expects for a layered superconductor in a magnetic field that there is superconducting phase order in all three dimensions or in none.) The condition of zero resistivity for currents parallel to the planes is maintained to much higher fields and temperatures than found for comparably doped LSCO \cite{sasa00,gila05}.  Thus, it appears that the special crystal structure of LBCO is capable of pinning the magnetic vortices via the induced stripe order, limiting dissipative flux flow, though not preventing slips in the phase of the superconducting order parameter when currents flow along the $c$ axis.

\section{Acknowledgements}
I am grateful to my numerous collaborators, and especially to Steve Kivelson and Eduardo Fradkin for many stimulating and illuminating conversations.
This work was supported by the Office of Basic Energy Sciences, Division of Materials Science and Engineering, U.S. Department of Energy (DOE), under Contract No. DE-AC02-98CH10886 and through the Center for Emergent Superconductivity, an Energy Frontier Research Center. 


\end{document}